\documentclass[11pt, twoside]{article}
\usepackage{hyperref}
\usepackage[a4paper, left=.5in, right=.5in, top=1in, bottom=1in]{geometry} 
\usepackage{fancyhdr} 
\usepackage{multicol}  
\usepackage{graphicx}  
\usepackage{amsmath}   
\usepackage{amssymb}   
\usepackage{graphicx}  
\usepackage{dcolumn}  
\usepackage{bm}  
\usepackage[sort&compress]{natbib}
\usepackage{amssymb}
\usepackage{multirow}
\usepackage{amsmath}
\usepackage{cases}
\usepackage{placeins}
\usepackage{xcolor}
\usepackage{hyperref}
\usepackage[figurename=Figure]{caption}
\usepackage{caption}
\usepackage{subcaption}
\usepackage{booktabs} 
\usepackage[utf8]{inputenc}
\usepackage{chemformula}
\usepackage{soul}
\usepackage{easyReview}
\usepackage{todonotes}
\usepackage{breqn}
\usepackage{amsmath}
\usepackage{booktabs}
\usepackage{array}
\usepackage{longtable}
\usepackage{lscape}
\usepackage[version=3]{mhchem} 
\usepackage{lipsum}
\usepackage{authblk}
\usepackage{url}
\makeatletter
\newcommand\ntsty{\@setfontsize\ntsty{6.31415}{7.1828}}
\newcommand\fontbigt{\@setfontsize\notsotiny{7}{6}}
\makeatother
\makeatletter
\newcommand\fontsmallt{\@setfontsize\notsotiny{7}{8}}
\makeatother
\begin{document}

\title{Optoelectronic properties and performance optimization for photovoltaic applications of R3m-RbGeX$_3$ (X = Cl, Br, I)
perovskites: A combined DFT and SCAPS-1D study}
\author[1]{Piyush Kumar Dash}
\author[1]{Palash Banarjee}
\author[1,2]{Anupriya Nyayban}
\author[1]{Subhasis Panda\footnote{corresponding author, email : subhasis@phy.nits.ac.in}}

\affil[1]{Department of Physics, National Institute of Technology Silchar, Assam 788010, India}
\affil[2]{Department of Physical Sciences, Indian Institute of Science Education and
Research Kolkata, Mohanpur, W.B. 741246, India }

\date{}
\maketitle

\section*{Abstract}

In pursuit of an all-inorganic non-toxic perovskite solar cell (PSC) with enhanced performance, we have investigated the rhombohedral phase of the germanium-based rubidium halide perovskites RbGeX$_3$ (X = Cl, Br, I). The structural analysis followed by an in-depth study of the electronic and optical properties of these materials is performed within the framework of density functional theory (DFT). A detailed investigation of the electronic properties is carried out by examining the band structure and partial density of states (PDOS). PBE and TB-mBJ exchange-correlation functionals are used with and without the spin-orbit coupling (SOC), thereby obtaining accurate predictions of the band gaps. The key optical properties such as the real and imaginary parts of the dielectric function, absorption coefficient and refractive index are studied using PBE functional and compared among the three halides. RbGeI$_3$ exhibits the lowest band gap of 0.96 eV with the TB-mBJ + SOC functional, along with the most favorable optical properties, hence, it was identified as the most suitable candidate for the absorber layer (AL) of the PSC. SCAPS-1D simulation is performed using various input parameters for the AL such as the band gap, the effective densities of states for the conduction and valence bands and the electron and hole mobilities extracted from the DFT calculation. Performance optimization is done by exploring the impact of different inorganic hole transport layers (HTLs) and electron transport layers (ETLs). The impact of different layer thicknesses, doping densities, defect densities at the AL, ETL/AL, and AL/HTL interfaces, various back contacts, and the influence of series and shunt resistance on the overall device performance are studied. The optimized all-inorganic device demonstrated a remarkable power conversion efficiency (PCE) of $25.76\%$ with a fill factor (FF) of $79.81\%$ for the configuration of FTO/SnO$_2$/RbGeI$_3$/CuI/Au.
\newpage
\section{Introduction}
Research into perovskite materials has recently surged due to the potential of affordable solution-processed perovskite solar cells with an efficiency exceeding 20\% \cite{NREL_CellEfficiency}. These materials display a variety of fascinating optoelectronic properties, making them suitable candidates for various photovoltaic applications. However, their structural and mechanical attributes present significant scientific challenges \cite{article}. The most commonly studied perovskites with high efficiency are the hybrid organic-inorganic perovskites, following the formula ABX$_3$, where A represents an organic cation, X signifies a halide anion and B is a group VI heavy metal cation such as lead. Hybrid lead halides have gained considerable attention in solar cell applications due to their high absorption, easy synthesis, adjustable band gaps, long carrier diffusion lengths and compatibility with solution processing \cite{wang2015high, PhysRevApplied.2.034007}. Their initial application in photovoltaic cells is reported by Kojima et al. in 2009, with an efficiency of 3.8\% \cite{Kojima2009}. Over the last decade, the power conversion efficiency has reached up to 25.5\% \cite{wu, Yoo}. However, the toxicity of lead in these perovskites remains a significant obstacle for large-scale commercialization. 
 
 In view of that researchers have been investigating alternative divalent metal cations like Sn$^{2+}$ and Ge$^{2+}$, which share a +2 oxidation state and similar electronic configurations with Pb$^{2+}$ \cite{3}. Sn$^{2+}$ has an ionic radius of 1.35 Å, which is smaller than that of Pb$^{2+}$. This allows Sn-based perovskites to maintain structural stability when Pb$^{2+}$ is substituted \cite{Noel2014}. These materials can potentially achieve higher theoretical efficiencies due to their smaller band gaps \cite{Roknuzzaman2017}. Sabba et al. examined the band gaps of Sn based halide perovskites and studied their halide variation \cite{Sabba2015}. CsSnI$_3$ is considered to be most suitable for PSC, with strong optoelectronic properties and a theoretically predicted PCE of 23\% \cite{Kumar2014, Wu2017}  and experimental PCE of 12.96\% \cite{chen2016leadfree}. However, Sn$^{2+}$ tends to oxidize to Sn$^{4+}$, leading to rapid degradation of Sn-based PSCs, severely limiting their efficiency \cite{Wei2021}. The low energy required for Sn vacancy formation and the tendency for oxidation from Sn$^{2+}$ to Sn$^{4+}$ result in self-doping and p-type metallic behavior \cite{Wei2021}.

Ge$^{2+}$ is another candidate to replace Pb$^{2+}$ with an even smaller ionic radius of 0.73 Å compared to Sn$^{2+}$. Inorganic Ge based perovskites exhibit greater ionic conductivity and possess wider band gaps than their Pb or Sn based counterparts. Raj et al. simulated CsGeI$_3$ PSCs using SCAPS-1D, achieving a peak PCE of 18.3\% and FF = 75.46\% \cite{Raj1}. Krishnamoorthy et al. were the first to fabricate CsGeI$_3$ based PSCs, achieving a poor PCE of 0.11\% \cite{Krishnamoorthy}. Chen et al. later synthesized high-quality CsGeX$_3$ quantum rods via a solvothermal process, achieving PCEs of 4.94\%, 4.92\% and 2.57\% for X = I, Br and Cl  respectively \cite{C8RA01150H}. The challenging synthesis process and poor film-forming quality of CsGeX$_3$ is a vital factor in their poor performance \cite{Wei2021}. Quan et al. revealed that the crystal structure of CsGeI$_3$ undergoes a phase transition from the orthorhombic phase to the rhombohedral (R3m) phase as the temperature decreases. This transformation is attributed to the shift in the atomic positions of Ge and I within the lattice \cite{wu2018csgei3}.

 Rubidium is a notable inorganic element that can be utilized in photovoltaic applications as rubidium based halide perovskites show inherent stability against moisture and temperature variations. In 2021 Jayan et al. studied several characteristics including the optoelectronic properties of cubic RbGeI$_3$ perovskites using different exchange-correlation functionals \cite{DeepthiJayan2021}. Various theoretical modeling of  cubic RbGeI$_3$ is performed using SCAPS-1D simulation software and predicted efficiencies of 10.11\%  in 2022 \cite{se} and of 17.93\% in 2024 \cite{talukdar2024rbgei3}. In 1989 Thiele et al. performed an experimental study on the rhombohedral phase of RbGeI$_3$ and found the lattice constant to be 5.99 Å \cite{Thiele1989DieKU}. With no further experimental study on the rhombohedral phase of RbGeX$_3$, in 2019 Jong et al. performed the first DFT study calculating the band gaps using the PBEsol functional \cite{ugjong}. As for the isostructural CsGeI$_3$, PBEsol functional is not reliable for the prediction of the band gap, in this paper we have explored different functionals to attain better prediction. No theoretical modeling of the PSC has been done with the rhombohedral phase of RbGeI$_3$ as the absorber layer. The gap in the literature motivates this paper for a better optoelectronic study of R3m-RbGeX$_3$ and the theoretical modeling of the PSC.

In the theoretical modeling of the PSC the choice of the charge transport layers is very crucial. The instability of organic compounds in PSCs is a major challenge \cite{Ju2018,Leijtens}. The state-of-the-art hole transport material Spiro-OMeTAD requires a complex and costly synthesis involving multiple steps such as cyclization, bromination, Grignard reactions and Hartwig-Buchwald coupling \cite{Saragi2007, Tour1990, Abdellah2021}. Organic charge transport materials also degrade under illumination and ambient conditions, affecting stability \cite{Akbulatov2017, Jena2018,Norrman}. This has led to a shift towards inorganic charge transport materials (NiO,  TiO$_2$, Cu$_2$O, CuSCN, ZnO, WS$_2$, CuI etc) which offer simpler fabrication processes, lower costs, better chemical and thermal stability, wider band gaps, higher carrier mobility, and greater transparency across UV, visible and IR spectra \cite{Liang2016, Ouedraogo2020, Singh2019, Wei2021}.

In the first part of this study, we have investigated the rhombohedral phase of RbGeX$_3$ (R3m space group) using DFT. Section 2 delineates details of the computational technique used in the study.  The structural properties are discussed in section 3. The electronic and optical properties are analyzed in section 4 and 5, respectively. Section 6 provides an initial configuration for the device simulation using SCAPS-1D. Section 7 focuses on the analysis of performance optimization of the R3m-RbGeI$_3$ based PSC by using various inorganic HTLs and ETLs and changing their thicknesses, absorber layer thickness, doping densities for AL, HTL and ETL, defect densities at interfaces, back contacts, series and shunt resistances. Finally, we have achieved a PCE of 25.76\% and FF of 79.81\% with a concluding remark in section 8.

\section{Computational details for DFT analysis}
 The computational investigations in this study are performed using the WIEN2k software package that implements the full-potential linearized augmented plane wave (FP-LAPW) method \cite{blaha}. For the exchange-correlation potential, we employed two commonly used functionals: the Perdew, Burke, and Ernzerhof (PBE) generalized gradient approximation (GGA) and the Tran-Blaha-modified Becke-Johnson (TB-mBJ) potential with and without spin-orbit coupling (SOC) effect. The PBE functional is widely adopted due to its favorable balance between computational efficiency and accuracy while the TB-mBJ functional provides a more accurate approximation of the band gap compared to PBE. TB-mBJ provides results that align better with the experimental results. Its accuracy is comparable to hybrid functionals or many-body GW methods, which are computationally expensive \cite{PhysRevLett.102.226401}. Given these advantages, band gap calculations in this study are performed using the TB-mBJ potential where as optical properties are computed using the PBE functional.

 The parameter $RK_{max}$, which controls the size of the basis sets in the FP-LAPW method, is set to a value of 9 for all three halide perovskites. For all rubidium (Rb) atoms, a muffin-tin radius of 2.5 atomic units (a.u.) is assigned, while a radius of 2.2 a.u. is used for germanium (Ge) atoms. The muffin-tin radii for the halogen atoms (Cl, Br, I) are set to 1.99, 2.17 and 2.36 a.u. respectively, based on their atomic sizes. Brillouin zone integrations are performed using the tetrahedron method. For the electronic structure calculations, a k-point mesh of $10 \times 10 \times 10$ is employed, which provides a sufficiently dense sampling for the required accuracy. The convergence criteria for the energy, charge and force are set to be $1 \times 10^{-5}$ Ry, $1 \times 10^{-5}$ e, and 1 mRy/a.u., respectively, ensuring that the results are well-converged. The lattice parameters of each material are determined by fitting the energy-volume data to the Birch-Murnaghan equation of state (EOS).
 For the calculation of optical properties, the OPTIC module integrated within WIEN2k \cite{optic} is used. This module enables the calculation of real and imaginary parts of the  dielectric function, refractive indices and absorption spectra based on the electronic structure. To improve the accuracy of the optical properties calculations, a denser k-point mesh of $21 \times 21 \times 21$ is employed, which reduces to 891 k-points in the irreducible Brillouin zone (IBZ).

\begin{figure}
\hspace{.5cm}
\begin{subfigure}{0.45\textwidth}
\includegraphics[width=\textwidth]{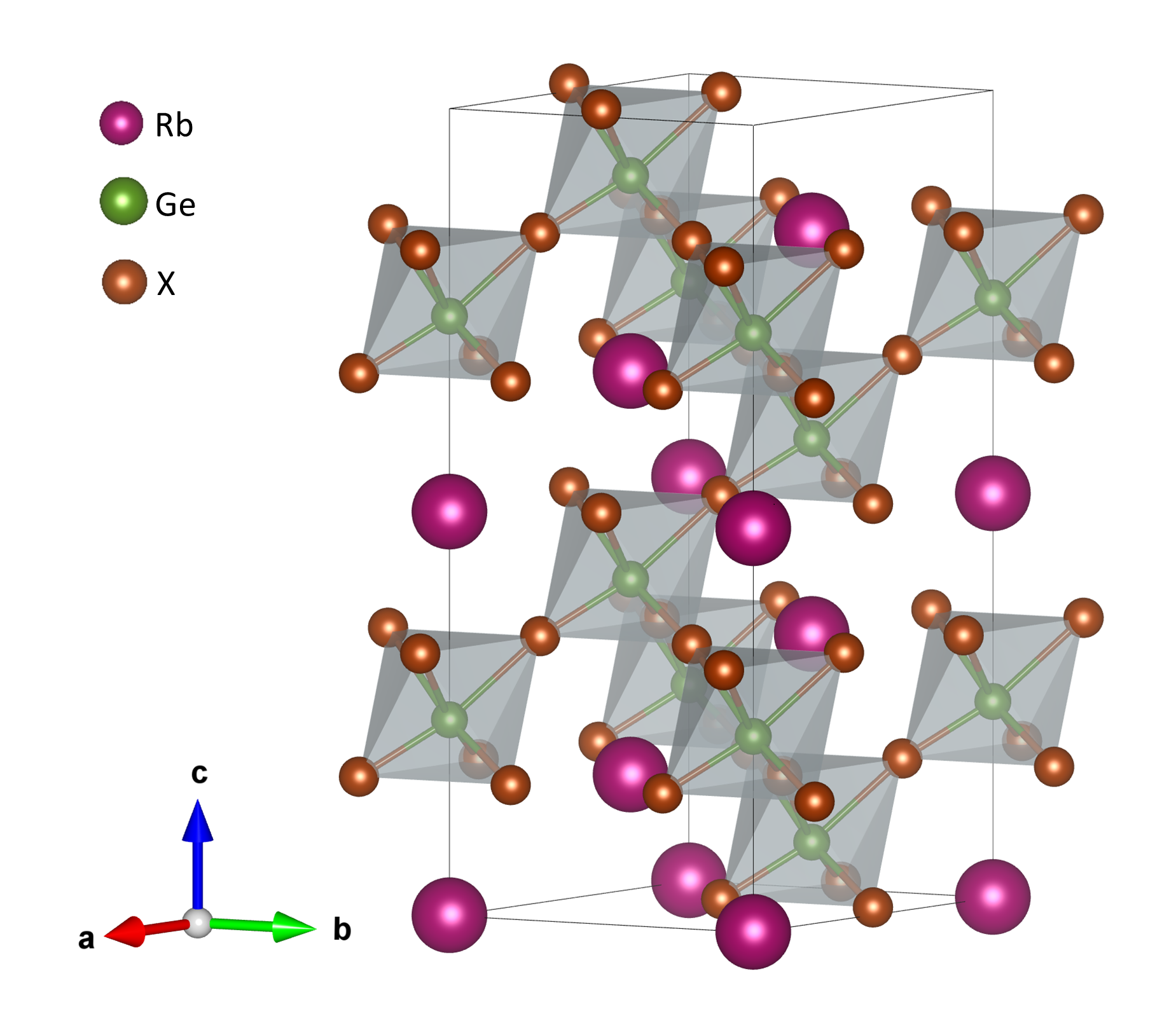}
\caption{}
\label{fig:(a)}
\end{subfigure}
\begin{subfigure}{0.5\textwidth}
\includegraphics[width=\textwidth]{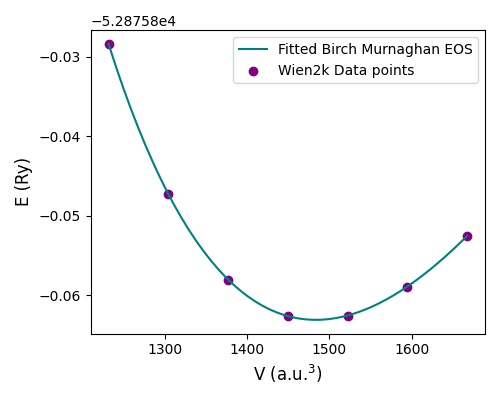}
\caption{}
\label{fig:(b)}
\end{subfigure}
\caption{(a) Crystal structure of RbGeX$_3$  (b) Volume optimization curve - total energy (E) vs volume (V) for RbGeI$_3$. }
 \end{figure}

\section{Structural analysis} 

This study investigates the rhombohedral phase of three materials RbGeCl$_3$, RbGeBr$_3$ and RbGeI$_3$, all of which crystallize in the space group R3m (SG-160). In this phase, each unit cell contains one formula unit, with four halide (X) atoms occupying the equatorial positions and two X atoms located at the apical positions. Rubidium atoms are situated at the voids of the GeX$_6$ octahedra, as shown in figure \ref{fig:(a)}. The apical and equatorial Ge-X bond lengths are calculated to be 3.04 Å, 2.85 Å and 2.72 Å for X = I, Cl, Br respectively. The decrease in the bond lengths from I to Cl to Br is due to the decrease in ionic radius of X atoms. Structural optimizations revealed that the corner-sharing GeX$_6$ octahedra experience subtle distortions due to the off-centering of Ge$^{2+}$  cations, a behavior also observed in the cubic phases of CsSnX$_3$ and CsPbX$_3$ \cite{Yang2017}. This distortion induces a spontaneous electric polarization within the material resulting from the asymmetric displacement of the Ge$^{2+}$ ions. The presence of this polarization facilitates the enhanced separation of charge carriers that results in improving the performance of optoelectronic devices. The induced polarization can lead to a photo-voltage exceeding the band gap of the material \cite{Leguy2015}.

The stability of these perovskite structures is related to the Goldschmidt tolerance factor, which is given by the following expression:
\begin{equation*}
  t_G = \frac{r_A + r_X}{\sqrt{2} (r_B + r_X)},  
\end{equation*}
\noindent where $r_A$, $r_B$ and $r_X$ are the Shannon radii of the A, B and X ions, respectively. The calculated tolerance factors are listed in table \ref{tab:st} which are in agreement with the existing literature \cite{ugjong}. These values fall within the range (0.8 to 1.0) typically associated with stable perovskite structures. The optimized lattice parameters and lattice angles for all three materials are listed in table \ref{tab:st}. As the ionic radius of the X-site anion decreases from I to Cl the lattice parameter is seen to be decreasing, indicating a contraction of the unit cell. The calculated lattice constant for RbGeI$_3$ is in good agreement with the experimentally found value by Thiele et al. \cite{Thiele1989DieKU}.

\begin{table}[h]
    \centering

    \begin{tabular}{@{}lccccccc@{}}
        \toprule
         Material & \multicolumn{2}{c}{Tolerance factor $t_G$} & \multicolumn{1}{c}{Lattice angle } & \multicolumn{3}{c}{Lattice parameter (Å)} \\
       \cmidrule{2-7}
       & our cal. & other cal.\cite{ugjong} & $\alpha$ (degree) & our cal.& other cal.\cite{ugjong} & expt. \cite{Thiele1989DieKU}\\
       \midrule
       RbGeI$_3$ & 0.89 & 0.90 & 88.34  & 6.04 & 5.91 & 5.99  \\
       RbGeBr$_3$ & 0.91 & 0.92 & 87.73 & 5.68 & 5.53 & - \\
       RbGeCl$_3$ & 0.93 & 0.93 & 89.18 & 5.44 & 5.27 & - \\
       \bottomrule
       
       \end{tabular}

\caption{Comparison of tolerance factors, lattice angles and lattice parameters of RbGeX$_3$.}
\label{tab:st}
\end{table}
\section{Electronic properties}

\begin{figure}[h!]
    \centering

    \begin{subfigure}[b]{0.45\textwidth}
        \includegraphics[width=\textwidth]{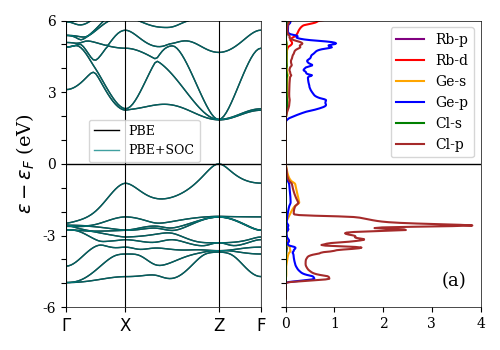}
    
    \end{subfigure}
    \begin{subfigure}[b]{0.45\textwidth}
        \includegraphics[width=\textwidth]{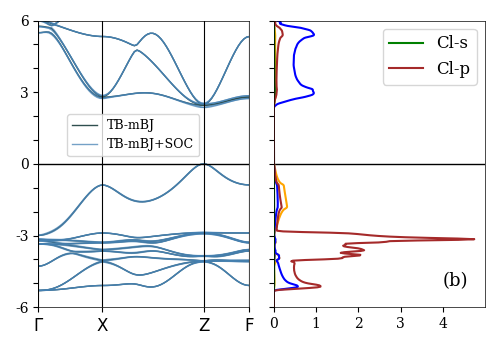}
  
    \end{subfigure}


    \begin{subfigure}[b]{0.45\textwidth}
        \includegraphics[width=\textwidth]{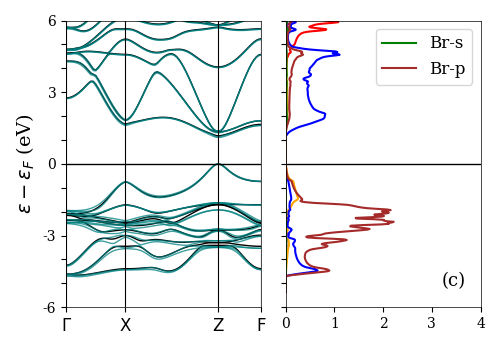}

    \end{subfigure}
    \begin{subfigure}[b]{0.45\textwidth}
        \includegraphics[width=\textwidth]{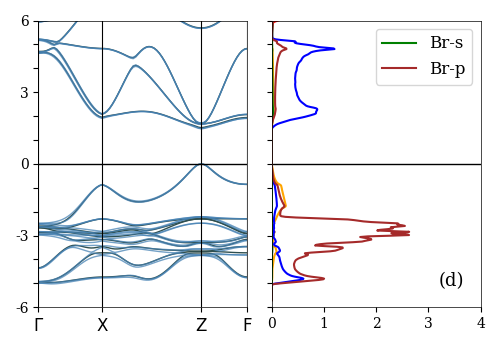}

    \end{subfigure}


    \begin{subfigure}[b]{0.45\textwidth}
        \includegraphics[width=\textwidth]{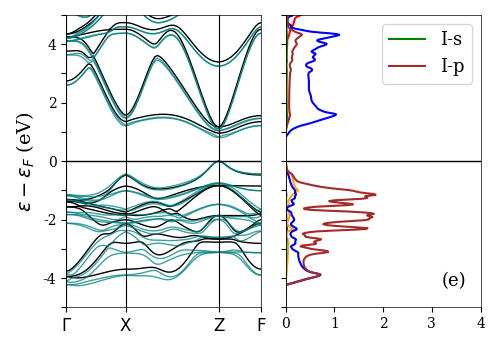}

    \end{subfigure}
    \begin{subfigure}[b]{0.45\textwidth}
        \includegraphics[width=\textwidth]{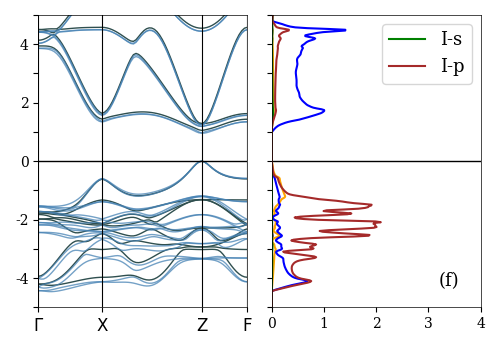}
     
    \end{subfigure}

    \caption{Band structures with and without SOC and PDOS with SOC calculated using PBE (left column) and TB-mBJ (right column) functionals for RbGeCl$_3$: ((a), (b)), RbGeBr$_3$: ((c), (d)) and RbGeI$_3$: ((e), (f)).}
    \label{fig:bandstructures}
\end{figure}

 The electronic properties of the RbGeX$_3$ (X = Cl, Br, I) compounds are critically analyzed through their electronic band structures and PDOS. A key feature observed from the band structure calculations is that both the conduction band minimum (CBM) and the valence band maximum (VBM) occur at the Z high-symmetry point, signifying the direct band gap nature of RbGeX$_3$. This nature is highly desirable for optoelectronic applications as it allows for efficient light absorption and emission. The band structure calculations are performed using the PBE-GGA and TB-mBJ potentials, with and without SOC and PDOS with SOC for all three materials and and are presented in figure \ref{fig:bandstructures} (a-f).  

A comparison of the calculated band gaps for isostructural CsGeCl$_3$, using PBE, PBEsol, HSE06 and TB-mBJ potentials with and without SOC in table \ref{bandgapscsgecl3} reveals that the TB-mBJ potential with SOC provides the most accurate prediction, in line with experimental results. To the best of our knowledge, there is no reported experimental data for of RbGeX$_3$ in R3m phase, hence the TB-mBJ + SOC method is adopted as it is the most reliable for estimating the band gap. The calculated band gaps for RbGeX$_3$ are summarized in table \ref{bandgap}. The band gap shows a consistent trend decreasing from Cl to I, which aligns with earlier studies on both RbGeX$_3$ \cite{ugjong} and CsGeX$_3$ \cite{nguyen, ELAKKEL2024112903}. This behavior indicates the impact of halide substitution on the electronic structure and the band gap, suggesting that larger halides lead to a reduction in the band gap. 

The PDOS analysis further elucidates the contributions of various orbitals to the electronic states near the Fermi level. From the figure \ref{fig:bandstructures} it is evident that for the conduction band in all three halides, the dominant contributions come from the p orbital of Ge whereas for the valance band, 4s orbital of Ge and p orbital of X atom play significant roles. 4s orbital of Ge is playing a notable role for RbGeCl$_3$ along with a contribution from p orbital of the halide atom, moving from Cl to I the p orbital of X overtakes and becomes dominant. The energy states below the Fermi level, as calculated with PBE+SOC, range from 0 to -4.04 eV for RbGeI$_3$, 0 to -4.6 eV for RbGeBr$_3$, and 0 to -4.9 eV for RbGeCl$_3$, with the p orbitals of the halide atoms being the primary contributors. For the TB-mBJ+SOC calculations, the energy states below the Fermi level extend from 0 to -4.4 eV for RbGeI$_3$, 0 to -5.07 eV for RbGeBr$_3$, and 0 to -5.3 eV for RbGeCl$_3$, with the halide p orbitals still being the dominant contributor. These results confirm the important role of halide orbitals in shaping the electronic structure of these materials. Above the Fermi level, the broad peaks for the CBM extend from 0.82 to 6.1 eV for RbGeI$_3$, 1.11 to 5.1 eV for RbGeBr$_3$, and 1.74 to 5.6 eV for RbGeCl$_3$, as observed in the PBE+SOC calculations. These energy ranges correspond to the conduction states, where the Ge 4p orbitals dominate the contributions. In the TB-mBJ+SOC calculations, the first broad peak spans from 0.96 to 6.2 eV for RbGeI$_3$, 1.47 to 5.29 eV for RbGeBr$_3$, and 2.36 to 6.1 eV for RbGeCl$_3$. The contributions of orbitals to both VBM and CBM are similar in both the PBE+SOC and TB-mBJ+SOC calculations, which also aligns with the patterns seen in PDOS of isostructural R3m CsGeI$_3$ \cite{qui}. The observed trends in the band gaps and the orbital contributions to the electronic states highlight the critical role of the halide composition for the properties of these materials and their photovoltaic application.

\begin{table}[h] 
\begin{footnotesize}
    \centering
    \begin{tabular}{@{}lcccccccccc@{}}        \toprule
        & \multicolumn{4}{c}{Without SOC} & \multicolumn{4}{c}{With SOC}  & \multirow{2}{*}{Exp\cite{nguyen}} \\ \cmidrule(r){2-5}\cmidrule(lr){6-9}
        & PBE\cite{nguyen} & PBEsol\cite{ugjong} & HSE06\cite{nguyen} & TB-mBJ\cite{ELAKKEL2024112903} & PBE\cite{nguyen} & PBEsol\cite{ugjong} & HSE06\cite{ugjong} & TB-mBJ\cite{ELAKKEL2024112903} & & \\
        \midrule
        CsGeCl$_3$ & 2.12 & 2.13 & 2.81 & 3.63 & 2.05 & 2.08 & 2.28 & 3.61 &  3.40 \\
        \bottomrule
    \end{tabular}
    \caption{Band gaps in (eV) for CsGeCl\textsubscript{3} using different functionals.}
    \label{bandgapscsgecl3}
    \end{footnotesize}
\end{table}

\begin{table}[h]
    \centering

    \begin{tabular}{@{}lcccccccc@{}}
        \toprule
        Material & PBE & PBE + SOC & TB-mBJ & TB-mBJ + SOC \\
        \midrule
        
        RbGeCl\textsubscript{3} & 1.84 & 1.74 & 2.45 & 2.36 \\
        
        RbGeBr\textsubscript{3} & 1.17 & 1.11 & 1.51 & 1.47 \\
        
        RbGeI\textsubscript{3} & 0.95 & 0.82 & 1.05 & 0.96 \\
        
        \bottomrule
    \end{tabular}
          \caption{Band gaps in (eV) for RbGeX$_3$ using PBE and TB-mBJ functionlals with and without SOC.}
    \label{bandgap}

\end{table}

\section{Optical properties}

The optical properties of the RbGeX$_3$ (X = Cl, Br, I) perovskite halides are systematically examined and compared to assess their suitability and the superiority of RbGeI$_3$ for photovoltaic applications. Key optical parameters, including the real and imaginary components of the dielectric function, absorption coefficient and refractive index are computed and analyzed as a function of incident photon energy to gain deeper insights into their optoelectronic characteristics. Notably, the optical properties along the $xx$ and $yy$ directions are found to be identical, reflecting the symmetry inherent in the rhombohedral crystal structure.

\noindent \textbf{Dielectric function $\varepsilon(\omega)$-}

The optical properties of halide perovskites are fundamentally described by their complex dielectric function $\varepsilon(\omega)$, which is a crucial parameter for understanding how these materials interact with electromagnetic radiation. The complex dielectric function is expressed as $\varepsilon(\omega)$ = $\varepsilon_1(\omega) + i \varepsilon_2(\omega)$. In this expression, {$\varepsilon_1(\omega)$} is the real part of the dielectric function which reflects the dispersion of the material. {$\varepsilon_2(\omega)$}, the imaginary part, corresponds to the absorption characteristics of the material, indicating how effectively the material absorbs light at different frequencies.

\begin{figure}
    \centering
    \includegraphics[scale=.9]{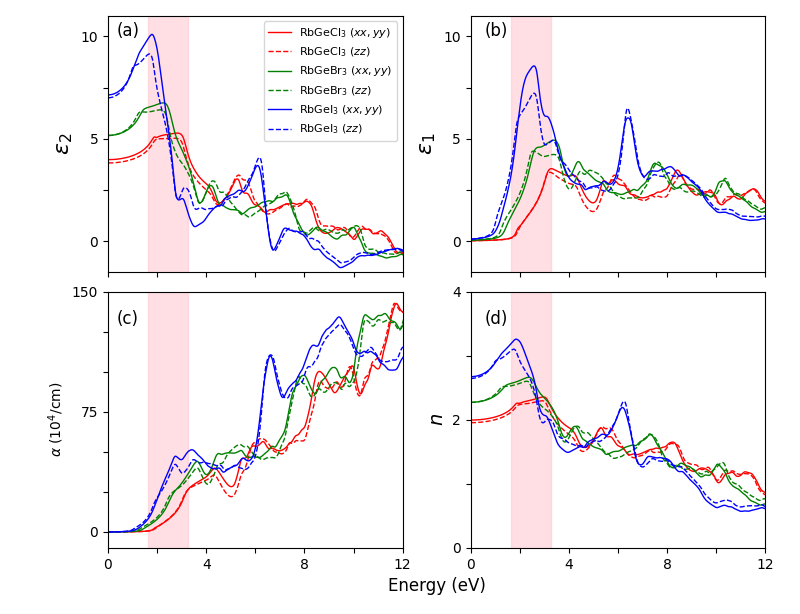}
    \caption{ Variation of the imaginary ($\varepsilon_2$) and the real ($\varepsilon_1$) part of the dielectric function (a,b), absorption coefficient ($\alpha$) (c), refractive index ($n$) (d) with respect to the incident photon energy along all three directions.}
    \label{opt}
\end{figure}
For semiconductors like RbGeX$_3$, the analysis often neglects intraband transitions due to their minimal impact on optical properties in comparison to interband transitions. Interband transitions, which involve electron transitions between different energy bands, are more significant and are categorized into direct and indirect types. Direct interband transitions are electron movements between the valence and conduction bands that occur without phonon involvement.  These transitions are crucial for determining the optical absorption spectrum of the material. To account for all possible transitions between occupied and unoccupied states, the calculation of the matrix elements involves integrating over the irreducible Brillouin zone, that is, $\varepsilon_2^{ss}= \dfrac{1}{N} \sum_{i=1}^{N} \sigma_i^T \varepsilon_2 (IBZ) \sigma_i,$ where $\sigma_i$ is a symmetry operation and $N$ is the number of symmetry operations.  Indirect interband transitions involve phonons to conserve momentum during the electron transition between bands. These transitions contribute less to the optical response at lower energies and are typically excluded from detailed optical analysis. The real part of the dielectric function, $\varepsilon_1(\omega)$, is obtained from $\varepsilon_2(\omega)$ using the following Kramers-Kronig relation - $
\varepsilon_1(\omega) = 1 + \dfrac{2}{\pi} \int_0^\infty \dfrac{\omega^{\prime} \varepsilon_2(\omega^{\prime})}{\omega^{\prime^2} - \omega^2} d\omega^{\prime}. $

\begin{table}[h!]
\centering
\begin{tabular}{lcccccc}
\toprule
Material & \multicolumn{2}{c}{RbGeCl$_3$} & \multicolumn{2}{c}{RbGeBr$_3$} & \multicolumn{2}{c}{RbGeI$_3$} \\
\cmidrule(lr){2-3} \cmidrule(lr){4-5} \cmidrule(lr){6-7}
Direction & (100) & (001) & (100) & (001) & (100) & (001) \\
\midrule
$\varepsilon_1(0)$ & 3.98 & 3.82 & 5.16 & 5.15 & 7.14 & 6.99 \\
$n(0)$ & 1.99 & 1.95 & 2.27 & 2.27 & 2.67 & 2.64 \\
\bottomrule
\end{tabular}
\caption{Static values for $\varepsilon_1$ and $n$ for RbGeX$_3$ along different polarization directions.}
\label{tab:static_values_horizontal}
\end{table}

Figure \ref{opt}(a) presents the variation of the imaginary part of the dielectric function, $\varepsilon_2$, with incident photon energy for all three RbGeX$_3$ materials with pink shaded region indicating the visible range of the spectrum. The first peak is attributed to transitions involving mixed X-p states and Ge-4s states, transitioning to the 4p states of Ge in the conduction band. These peaks in $\varepsilon_2$ correspond to electronic transitions from the VBM to the CBM. The spectral shift towards higher photon energies as the halide changes from I to Cl can be attributed to the increase in the band gap. RbGeCl$_3$ shows the highest absorption onset around 3 eV indicating the largest band gap among the three. While RbGeI$_3$ exhibits the lowest absorption onset with distinct peaks at 2.51 eV with magnitudes of 8.55 for $xx$ and $yy$ direction and of 6.92 along $zz$ direction, signifying the smallest band gap and strongest absorption at lower energies.  RbGeCl$_3$ is thus more suitable for applications requiring UV-range absorption, whereas RbGeI$_3$ having the least band gap is useful for harvesting visible and near-infrared light. 

The variation of the real part of the dielectric function $\varepsilon_1$ is shown in figure \ref{opt}(b) with visible range represented as the pink shaded region.  Each graph indicates how strongly the material polarizes in response to an incoming electromagnetic field, with the peaks reflecting critical electronic transitions from the valence band to the conduction band. Initially, $\varepsilon_1$ increases with photon energy, reaching a peak before gradually decreasing. At certain photon energies, $\varepsilon_1$ drops below zero, suggesting metallic-like characteristics. For RbGeCl$_3$, the peak appears at slightly higher energies, reflecting its comparatively larger band gap and weaker response to lower-energy visible light. For RbGeBr$_3$ the peak shifts closer to the mid-visible range, resulting a dielectric response in that region. By contrast, RbGeI$_3$ exhibits the most pronounced peak at lower photon energies within the visible region, suggesting the least band gap among all three materials and allowing it to absorb and polarize strongly in the red and near-infrared ranges. The higher amplitude of $\epsilon_1$ for RbGeI$_3$ in this region shows its enhanced light–matter interaction and its potential for more efficient solar energy conversion. 

At zero photon energy, $\varepsilon_1(0)$ represents the static dielectric constant, which is observed to decrease in the order of RbGeI$_3$, RbGeBr$_3$ and RbGeCl$_3$, showing an inverse relationship with the material’s band gap. This trend is in line with previous studies on other such materials \cite{real1,real2}. The computed static dielectric constants for all the halide perovskites are listed in the table \ref{tab:static_values_horizontal}, where it is evident that RbGeI$_3$ shows the highest static dielectric constant value of 7.14 for $xx$ and $yy$ directions and 6.99 in $zz$ direction . A larger static dielectric constant in RbGeI$_3$ helps in screening the photo-generated carriers and suppress recombination losses hence improving the charge transport. 

\noindent \textbf{Absorption coefficient $\alpha(\omega)$-}

The absorption coefficient $\alpha(\omega)$, which describes how much light is absorbed when passed through the absorbing material per unit length is given by:
\begin{equation*}
\alpha(\omega) = \frac{\omega}{c} \varepsilon_2(\omega),
\end{equation*}

\noindent where $c$ represents the speed of light.  In figure \ref{opt}(c) , RbGeCl$_3$,  having the largest band gap among the three, shows a comparatively later onset of absorption and lower absorption coefficients in visible range. RbGeBr$_3$, having comparatively smaller band gap, begins absorbing at comparatively lower photon energies, producing a more pronounced rise in its absorption curve within the visible range. RbGeI$_3$ has the narrowest band gap and displays a strong and early absorption onset, maintaining a relatively high absorption coefficient throughout most of the visible range. RbGeI$_3$ exhibits a peak in the visible range and strong absorption in both the visible and ultraviolet regions, making it highly effective at capturing light across a broad spectrum. This characteristic is highly desirable for solar cells because it allows for the efficient conversion of a significant portion of solar radiation into electricity. In contrast, RbGeBr$_3$ and RbGeCl$_3$, absorb light having higher energy. Thus RbGeI$_3$ has an advantage in terms of light absorption, particularly in the energy range of the visible solar spectrum.

\textbf{\noindent Refractive index $n(\omega)$ -}

The refractive index $n(\omega)$ is an essential parameter providing information about how strongly the material bends or refracts incident light, which can be derived from $\varepsilon_1(\omega)$ and $\varepsilon_2(\omega)$ using the equation 

\begin{equation*}
n(\omega) = \sqrt{\frac{\varepsilon_1(\omega) + \sqrt{\varepsilon_1(\omega)^2 + \varepsilon_2(\omega)^2}}{2}}. 
\end{equation*}

Figure \ref{opt}(d) shows the variation of $n$ with photon energy. RbGeBr$_3$ and RbGeI$_3$ show peaks in the visible spectrum (pink shaded region) whereas RbGeCl$_3$ has first peak outside the visible spectrum indicating a poorer photovoltaic application. A shift of the peaks towards lower energy spectrum and their increasing value of the amplitude of the peak from Cl to I is observed which is associated with the decreasing values of band gap. The sharp peaks near 2 eV for RbGeI$_3$ indicates a superior light–matter interaction in lower energy visible range. The static refractive indices for all three materials are listed in table \ref{tab:static_values_horizontal}. RbGeI$_3$ exhibits a relatively high static refractive index of 2.67 along the (100) and (010) directions and 2.64 along the (001) direction, which means that light slows down more in RbGeI$_3$ compared to RbGeBr$_3$ and RbGeCl$_3$. This is advantageous for light trapping in photovoltaic devices, as the higher refractive index can enhance light absorption by increasing the likelihood of light interacting with the material. Materials with $n(0)$ values ranging from 2.0 to 4.0 are generally optimal for optoelectronic applications \cite{pccp} and both RbGeI$_3$ and RbGeBr$_3$ fall within this range.
From the overall optical property analysis of all the key parameters, RbGeI$_3$ is considered to be the most promising for photovoltaic applications.

\section{Device simulation methodology }
 The solar cell capacitance simulator (SCAPS-1D) \cite{sc} is selected for the simulation
 due to its advantageous features, including ease of use, low computational cost and flexibility in simulating complex hetero structure systems. This tool efficiently models up to seven-layer systems with adjustable interfaces to explore material effects on device performance.  The SCAPS-1D simulator is based on three fundamental semiconductor equations: Poisson’s equation and the continuity equation for electron and hole. These equations form the core of the simulation framework enabling the accurate modeling of charge carrier dynamics, electrostatic potential distribution and current flow within the semiconductor device structure.
The Poisson's equation reads
\begin{align*}
\frac{\partial}{\partial x} \left( -\epsilon(x) \frac{\partial V}{\partial x} \right) =
   q ( p(x) - n(x) + N_{D^+}(x) - N_{A^-}(x)
    + p_t(x) - n_t(x)),   
\end{align*}
 where $\epsilon(x)$ is the permittivity of the material, which can vary with the position $x$. $V$ is the electrostatic potential. $p(x)$($n(x)$) is the hole (electron) density. $N_{D^+}(x)$ is the density of ionized donor atoms. $N_{A^-}(x)$ is the density of ionized acceptor atoms. $p_t(x)$ and $n_t(x)$ represent density of trapped holes and electrons and
 $q$ is the elementary charge.

The continuity equation for electrons/holes reads:
\begin{equation*}
    \frac{\partial (n/p)}{\partial t} = -\frac{1}{q} \frac{\partial J_{(n/p)}}{\partial x} + G_{(n/p)} - R_{(n/p)} ,
\end{equation*}
where $n$ ($p$) represents the density of electrons (holes). $J_{(n/p)}$ denotes the current density of electrons (holes), while $G_{(n/p)}$ indicates the rate of electron (hole) generation. $R_{(n/p)}$ refers to the rate of electron (hole) recombination.

\begin{figure}
    \centering
    \includegraphics[scale=0.5]{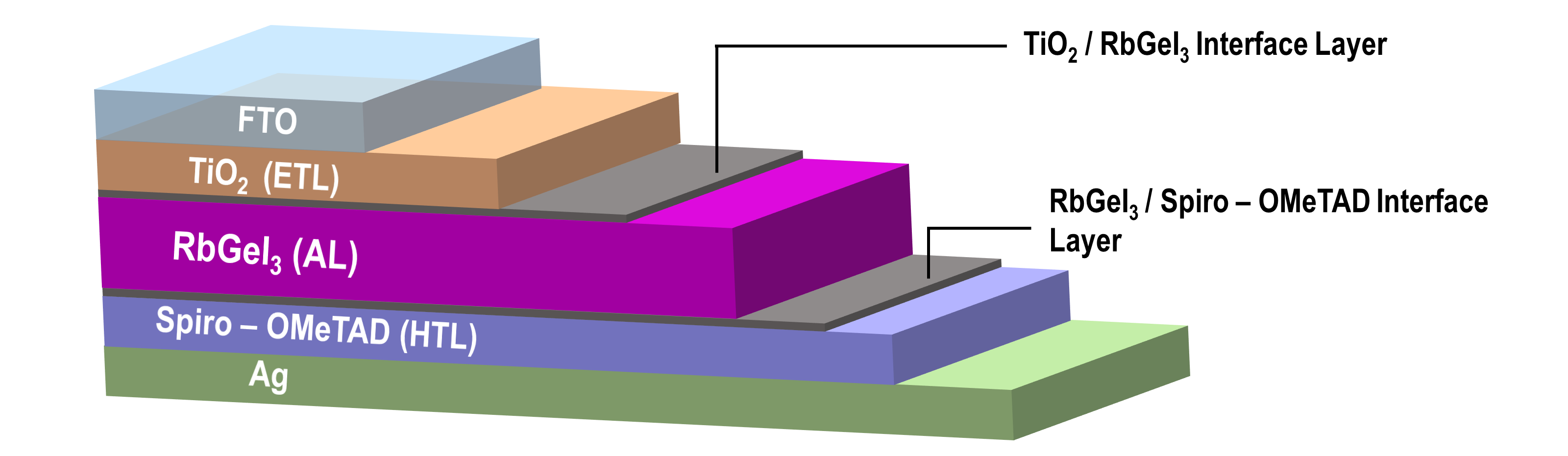}
    \caption{Initial configuration of the PSC.}
    \label{fig:initial}
\end{figure}

The initial simulation as illustrated in figure \ref{fig:initial} utilizes a multi-layered structure consisting of FTO-coated glass as the front contact, TiO$_2$ as the ETL, RbGeI$_3$ as the light-absorbing material, spiro-OMeTAD as the HTL and Ag as the back contact. Table \ref{big_table} outlines the key parameters used in this setup. These parameters include the band gap ($E_g$), the electron affinity ($\chi$), the relative permittivity ($\epsilon_r$), the effective density of states for the conduction and valence bands ($N_C$, $N_V$), the electron and hole mobilities ($\mu_e$, $\mu_p$), the donor and acceptor doping densities ($N_D$, $N_A$) and the defect density ($N_T$) of the materials.

\begin{table}[htbp]
\centering
\renewcommand{\arraystretch}{1.25}
\setlength{\tabcolsep}{2pt}
\ntsty{
\begin{tabular}{lcccccccccccccc}    \toprule
\multirow{2}{*}{Properties} & \multicolumn{4}{c}{Materials in initial configuration} & \multicolumn{4}{c}{Input parameters for different HTLs} & \multicolumn{4}{c}{Input parameters for various ETLs} \\  \cmidrule(lr){2-5} \cmidrule(lr){6-9} \cmidrule(lr){10-13}
& Sprio-OMeTAD\cite{ef} & RbGeI$_3$ & TiO$_2$\cite{ef} & FTO\cite{se} & NiO\cite{se} & Cu$_2$O\cite{se} & CuI\cite{se} & CuSCN\cite{se} & ZnO\cite{se} & SnO$_2$\cite{se} & CdS\cite{se} & WS$_2$\cite{se} \\    \midrule
Thickness (nm) & $200$ & $400$ & $30$ & $500$ & $50-700$ & $50-700$ & $50-700$ & $50-700$ & $10-200$ & $10-200$ & $10-200$  & $10-200$ \\
$E_g$ (eV) & $2.88$ & $0.963$ & $3.2$ & $3.2$ & $3.8$ & $2.17$ & $2.98$ & $3.4$ & $3.3$ & $3.5$ & $2.4$ & $1.87$  \\
$\chi$ (eV) & $2.05$ & $3.9$ & $4.0$ & $4.4$ & $1.4$ & $3$ & $2.1$ & $1.7$ & $4.1$ & $4$ & $4.18$ & $4.3$  \\
$\varepsilon_r$ & $3$ & $7$ & $100$ & $9$ & $10.7$ & $7.5$ & $6.5$ & $10$ &  $9$ & $9$ & $10$ & $11.9$  \\
$N_V$ (cm$^{-3}$) & $2.5 \times 10^{20}$ & $4.9 \times 10^{17}$ & $2 \times 10^{20}$ & $1.8 \times 10^{19}$ & $1 \times 10^{19}$ & $1.1 \times 10^{19}$ & $1 \times 10^{19}$ & $1.8 \times 10^{18}$  & $1.9 \times 10^{19}$ & $2.2 \times 10^{16}$ & $1.9 \times 10^{19}$ & $2.4 \times 10^{19}$ \\
$N_C$ (cm$^{-3}$) & $2.5 \times 10^{20}$ & $1.7 \times 10^{18}$ & $1 \times 10^{21}$ & $2.2 \times 10^{18}$ & $2.8 \times 10^{19}$ & $2 \times 10^{18}$ & $2.8 \times 10^{19}$ & $2.2 \times 10^{19}$ &  $2.2 \times 10^{18}$ & $2.2 \times 10^{17}$ & $2.2 \times 10^{18}$ & $1 \times 10^{19}$ \\
$\mu_e$ (cm$^{2}/$(V$\cdot$s)) & $0.0021$ & $103$ & $0.006$ & $20$ & $12$ & $200$ & $100$ & $100$ & $100$ & $200$ & $100$ & $260$ \\
$\mu_p$ (cm$^{2}/$(V$\cdot$s)) & $0.0026$ & $242$ & $0.006$ & $10$ & $28$ & $80$ & $43.9$ & $25$ & $25$ & $80$ & $43.9$ & $51$  \\
$N_D$ (cm$^{-3}$) & $0$ & $10^9$ & $5.06 \times 10^{19}$ & $10^{19}$ & $0$ & $0$ & $0$ & $0$  & $1 \times 10^{19}$ & $1 \times 10^{19}$ & $1 \times 10^{19}$ & $1 \times 10^{19}$ \\
$N_A$ (cm$^{-3}$) & $10^{18}$ & $10^9$ & $0$ & $0$ & $2 \times 10^{19}$ & $2 \times 10^{19}$ & $2 \times 10^{19}$ & $2 \times 10^{19}$  & $0$ & $0$ & $0$ & $0$ \\
$N_T$ (cm$^{-3}$) & $10^{15}$ & $10^{15}$ & $10^{15}$ & $10^{15}$ & $ 10^{15}$ & $10^{15}$ & $10^{15}$ & $10^{15}$ & $10^{15}$ & $10^{15}$ & $10^{15}$ & $10^{15}$  \\ \bottomrule
\end{tabular}}
\caption{Properties of various materials: initial configuration, HTLs and ETLs.}
\label{big_table}

\end{table}

From the DFT calculations we have extracted the key parameters of the absorber layer, RbGeI$_3$ for the SCAPS-1D simulation. The conduction band and valence band effective density of states are calculated to be $N_C$ = $1.7\times10^{18}$ $cm^{-1}$ and $N_V$ = $4.89\times10^{17}$ $cm^{-1}$ by the formulae 

\begin{equation*}
N_c = \frac{2 (2 \pi m_e^* k_B T)^{3/2}}{h^3}, \qquad
N_v = \frac{2 (2 \pi m_h^* k_B T)^{3/2}}{h^3},
\end{equation*}

\noindent where $T$ is taken to be 300K, and $m_e^*$ and $m_h^*$ are the effective mass of the electron and hole, which have been calculated from the band structure data and found to be 0.17m$_e$ and 0.07m$_e$ respectively. The electron and hole mobility values are calculated by using the formulae
\begin{equation*}
    \mu_e = \frac{e\tau}{m_e^*}, \qquad
    \mu_h = \frac{e\tau}{m_h^*} ,
\end{equation*}

\noindent where $\tau$ is the relaxation time taken to be $10^{-14}$ s which is prescribed value for perovskites and found to be 242 $cm^{2}V^{-1}s^{-1}$ and 103 $cm^{2}V^{-1}s^{-1}$ respectively.

Under standard illumination conditions of 1000 W/m$^{2}$ at AM 1.5 G and a temperature of 300 K, the thermal velocities of both electrons and holes are assumed to be $10^7$ cm/s for all the layers. The DFT calculated parameters such as band gap, valance and conduction band density of states, mobility of holes and electron are used as the input parameters for absorber layer. For a practical model of the perovskite solar cell we have incorporated interface layers of TiO${_2}$/RbGeI$_3$ and RbGeI$_3$/Spiro-OMeTAD. For both interfaces, a neutral defect density of $10^{16}$ cm$^{-2}$ with a characteristic energy of 0.1 eV is assumed. The back contact is taken as silver (Ag) having work function of 4.57 eV. $N_D$ and $N_A$ for RbGeI$_3$ are initially set to be $10^9$ cm$^{-3}$ and optimized in subsequent simulations. The calculated initial performance parameters for the PSC are as follows: open-circuit voltage ($V_{OC}$) = 0.527 V, short-circuit current density ($J_{SC}$) = 44.155 mA/cm$^2$, fill factor (FF) = 76.48\% and power conversion efficiency (PCE) = 17.83\%.

\section{SCAPS-1D results and discussion}
\subsection{HTL optimization}

\begin{figure}[h!]
    \centering
    \includegraphics[scale=.9]{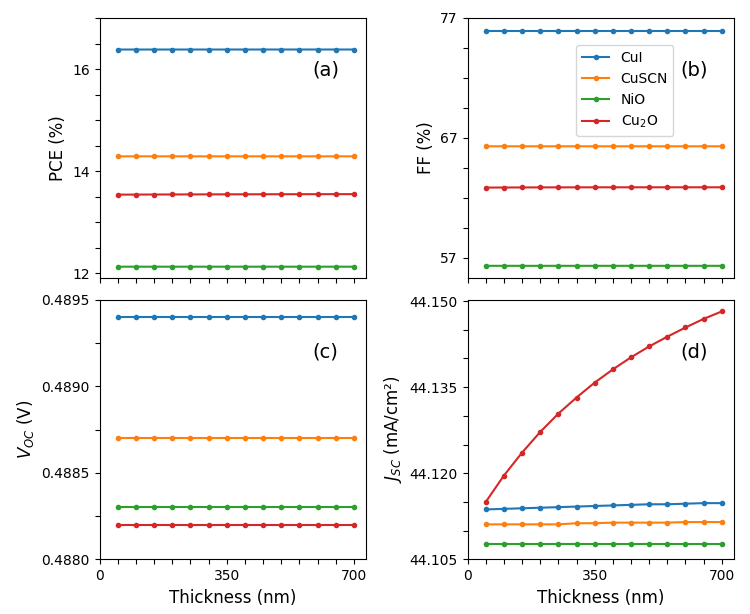}
 
\caption{Variation of (a) photo conversion efficiency (PCE) (b) fill factor (FF) (c) open circuit voltage ($V_{OC}$) and (d) short circuit current ($J_{SC}$) with respect to the thickness for different HTLs.}
\label{fightl}
\end{figure}

 The HTL plays a pivotal role in the performance of PSCs by facilitating hole transport, reducing electron-hole recombination, and providing a resistance to electron flow, thereby enabling the completion of the external circuit. Effective HTLs must possess high charge carrier mobility, appropriate doping levels, optimal thickness and a suitable band gap to ensure efficient device operation \cite{Saikia}. The performance of PSC is optimized by substituting the organic HTL Spiro-OMeTAD with stable inorganic HTLs (Cu$_2$O, CuSCN, CuI, NiO) having negative valance band offsets while maintaining TiO$_2$ as the ETL and $N_A$ to be  $2\times10^{19} $cm$^{-3}$. A positive value of the valence band offset creates a barrier that hinders the movement of photo-generated holes from the absorber to the HTLs. In contrast, when the VBO is negative, no such barrier exists, allowing photo-generated holes to flow smoothly towards the HTLs \cite{Ahmed2021}. The relevant input parameters for the different HTLs are summarized in table \ref{big_table}. To assess the influence of HTL thickness on the performance of the PSC, we varied the HTL thickness from 50 nm to 700 nm. The resulting device performance for different HTLs at various thicknesses is presented in figure \ref{fightl}. For all HTLs, the photovoltaic parameters exhibit only minor variations with increasing thickness, indicating the robustness of device performance. Specifically, $J_{SC}$ remains relatively constant for CuI, CuSCN and NiO, while Cu$_2$O shows a gradual increase in $J_{SC}$ with increasing thickness, suggesting improved charge collection. $V_{OC}$ remains stable across all HTLs with CuI exhibiting the highest value, indicating minimal recombination losses. CuI also demonstrates the highest PCE and fill factor maintaining superior device performance across the thickness range. The final performance parameters for all the HTLs with optimized thickness are listed in table \ref{htl_etl_performance}. Based on these findings, CuI with thickness 200 nm is selected as the optimal HTL for further simulations. The PCE of the PSC remains 16.38\%, with a fill factor of 75.87\%.

\subsection{ETL optimization}

\begin{figure}
    \centering    \includegraphics[scale=.9]{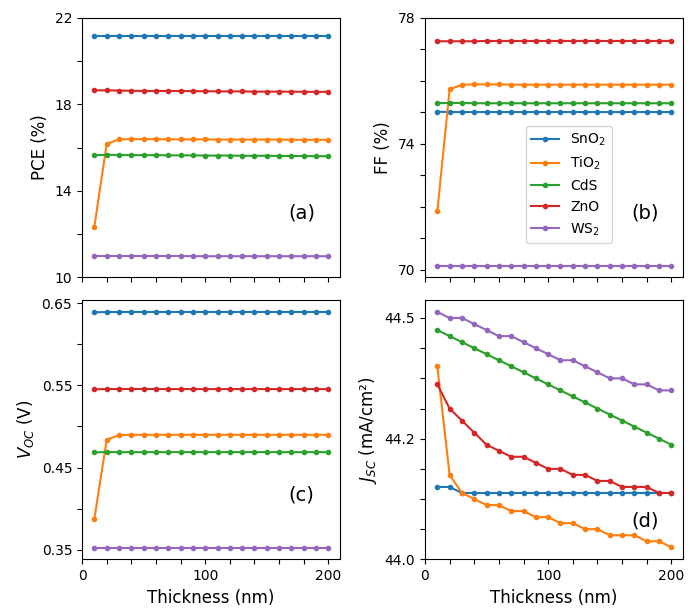}

\caption{Variation of (a) photo conversion efficiency (PCE) (b) fill factor (FF) (c) open circuit voltage ($V_{OC}$) and (d) short circuit current ($J_{SC}$) with respect to the thickness for different ETLs.}
\label{figetl}
\end{figure}

The role of the ETL is crucial in facilitating electron movement by obstructing the holes generated and reducing the recombination of holes with electrons. A good ETL should possess high electron mobility, an optimal thickness, appropriate doping levels and an adequate band gap. The ETL enables the extraction of photo-generated electrons from absorber layer, directing them to the front contact. The various inorganic ETLs such as ZnO, WS$_2$, CdS, SnO$_2$ are used to optimize the performance. The electron affinities of all the ETLs are greater than that of the AL, which is desirable for a smooth electron flow \cite{pccp}. From figure \ref{figetl}, the variation in performance parameters with different ETLs of varying thicknesses is demonstrated. The input parameters for these ETLs are provided in table \ref{big_table} where $N_D$ is taken to be $1 \times 10^{19}$ cm$^{-3}$ for all the materials for better comparison. The ETL thickness is varied from 10 nm to 200 nm and the performance parameters are analyzed. SnO$_2$ exhibits the highest efficiency around 21\% across the thickness range. TiO$_2$ initially shows lower efficiency at 10 nm  but stabilizes at around 16.38\% beyond 30 nm. The fill factor trends in figure \ref{figetl}(b) show that ZnO exhibits highest fill factors above 77\%, with minimal thickness dependence, while TiO$_2$ exhibits a sharp increase up 30 nm having a similar trend as its efficiency. In figure \ref{figetl}(c), $V_{OC}$ remains constant for most ETLs, with SnO$_2$ showing the highest values, indicating lower recombination losses. WS$_2$ consistently display lower $V_{OC}$, fill factor and efficiency values showing the poorest performance. As  shown in figure \ref{figetl}(d), $J_{SC}$ decreases with increasing thickness for all ETLs except SnO$_2$, which maintains an almost stable $J_{SC}$. The sharp decrease in the $J_{SC}$ of TiO$_2$ from 10 nm to 30 nm backs the trend seen in its efficiency and fill factor. Table \ref{htl_etl_performance} lists the performances for PSCs using various ETLs with optimal thicknesses. Based on these results, though ZnO exhibits the highest fill factor, with a modest fill factor and the highest efficiency, SnO$_2$ with a thickness of 30 nm is identified as the optimal ETL, providing an efficiency of 21.15\% and a fill factor of 75.1\%.

\begin{table}[ht]
\begin{footnotesize}

\centering

\setlength{\tabcolsep}{3pt}
\footnotesize{
\begin{tabular}{@{}ccccccc@{}}    \toprule
    PSC configuration & HTL variations &  ETL variations& $V_{OC}$ (V) & $J_{SC}$ (mA/cm$^2$) & FF (\%) & PCE (\%) \\ \midrule
    \multirow{4}{*}{FTO/TiO$_2$/RbGeI$_3$/HTL/Ag} & NiO & \multirow{4}{*}{--}  & 0.488 & 44.107 & 56.36 & 12.13 \\
   & CuI  &    & 0.489 & 44.113 & 75.87 & 16.38 \\
   & Cu$_2$O &  & 0.488 & 44.148 & 62.87 & 13.55 \\
   & CuSCN &   & 0.488 & 44.111 & 66.29 & 14.29 \\ \midrule
  \multirow{4}{*}{FTO/ETL/RbGeI$_3$/CuI/Ag} & \multirow{4}{*}{--} & SnO$_2$ & 0.639 & 44.120 & 75.01 & 21.15 \\
   & & CdS     & 0.468 & 44.366 & 75.29 & 15.65 \\
   & & WS$_2$  & 0.352 & 44.402 & 70.12 & 10.97 \\
   & & ZnO     & 0.545 & 44.231 & 77.26 & 18.64 \\ \bottomrule 
\end{tabular}}
\caption{Performance of PSCs with different HTLs and ETLs.}
\label{htl_etl_performance}
\end{footnotesize}
\end{table}

\subsection{AL thickness optimization}
\begin{figure}[h!]
   \centering    \includegraphics[scale=.9]{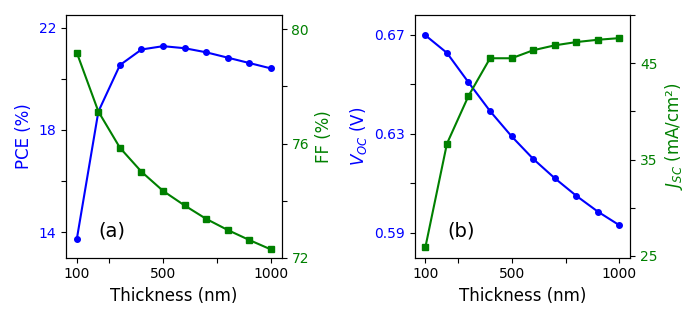}

    \caption{Variation of key photovoltaic parameters with respect to AL thickness: (a) Photo conversion efficiency (PCE) and fill factor (FF) (b) open circuit voltage ($V_{OC}$) and short circuit current ($J_{SC}$).}
    \label{al}
\end{figure}

The thickness of the perovskite layer impacts the diffusion lengths and lifetimes of photo-generated carriers \cite{Barbe2017,Kim2016}. We varied the RbGeI$_3$ layer thickness from 100 nm to 1000 nm to assess its effect on PSC performance as shown in figure \ref{al}. In figure \ref{al}(a), the efficiency shows a sharp increase from 13.72\% to 21.28\% as thickness increases from 100 nm to 500 nm. Beyond this point, the efficiency gradually declines. The fill factor follows a contrasting trend, decreasing consistently from around 79\% at 100 nm to approximately 72\% at 1000 nm. In figure \ref{al}(b), $V_{OC}$ decreases with increasing thickness, dropping from 0.67 V to 0.59 V. This decline suggests enhanced recombination and increased resistive losses \cite{Saikia}. On the other hand, $J_{SC}$ rises significantly with the thickness till 400 nm then shows a slight increase up to 1000 nm. The simultaneous reduction in $V_{OC}$ and FF ultimately results in lower overall performance at higher thicknesses hence an optimal absorber layer thickness 500 nm is considered to achieve a balance between the performance parameters having values : $V_{OC}$ = $0.629$ V, $J_{SC}$ = $45.51$ mA/cm$^2$, FF = $74.34$\% and PCE = $21.28$\%.

\subsection{Doping density optimization}

  \begin{figure}[h!]
       \centering
       \includegraphics[scale=1]{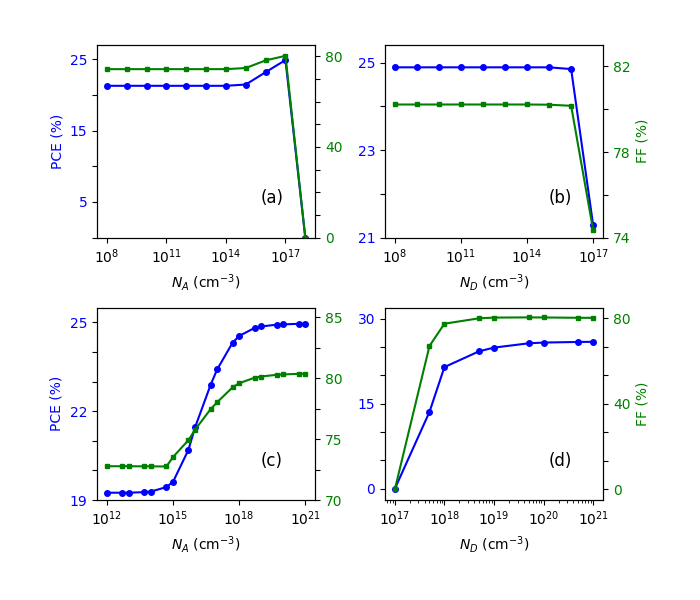}

    \caption{Variation of photo conversion efficiency (PCE) and fill factor (FF) with doping densities in different layers: (a) acceptor doping density ($N_A$) in AL, (b) donor doping density ($N_D$) in AL, (c) acceptor doping density ($N_A$) in HTL and (d) donor doping density ($N_D$) in ETL.}
    \label{doping}
\end{figure}

The performance of PSCs is strongly influenced by the doping densities within the perovskite layer. In the initial simulation, both the donor and acceptor doping densities in the perovskite layer were set to $10^9$ cm$^{-3}$. To explore the effect of doping density on PSC performance, we varied the acceptor concentration ($N_A$) in the perovskite layer from $10^8$ cm$^{-3}$ to $10^{18}$ cm$^{-3}$. Figure \ref{doping}(a) shows the variation of efficiency and fill factor with respect to $N_A$ of the AL. The efficiency and FF remain nearly constant up to $10^{15}$ cm$^{-3}$, after which both parameters increase, reaching their maximum at $10^{17}$ cm$^{-3}$. Beyond this, a sharp decline is observed due to increased charge carrier recombination \cite{Saikia}. Hence after taking the optimal $N_A$ as $10^{17}$ cm$^{-3}$ the effect of $N_D$ in the AL is analyzed and shown in figure \ref{doping}(b). The efficiency and FF remain exactly same up to $10^{16}$ cm$^{-3}$ and at $10^{17}$ cm$^{-3}$ both parameters decline sharply as it neutralizes the doping in the material. No donor doping is found to be necessary for optimal performance, with the optimal donor doping concentration $N_D = 0 $cm$^{-3}$ yielding equivalent device performance of FF = 80.21\% and PCE = 24.89\%.

Figure \ref{doping}(c) illustrates the impact of acceptor doping density in the HTL. A significant increase in both efficiency and FF is seen as $N_A$ rises from $5\times10^{12}$ cm$^{-3}$ to  $5\times10^{19}$ cm$^{-3}$  due to the enhanced internal electric field and charge transport velocity \cite{Ghosh2020}. Beyond that the parameters saturate, suggesting saturation of charge transport improvement. No significant change in performance is observed at higher doping concentrations. Figure \ref{doping}(d) evaluates the effect of donor doping density in the ETL. Both efficiency and FF exhibit a substantial increase up to $10^{17}$ cm$^{-3}$, after which they stabilize at $5\times10^{19}$ cm$^{-3}$, reflecting improved electron transport. However, excessive doping can lead to Coulomb traps, reducing electron mobility \cite{Saikia}. The optimal doping concentrations for HTL and ETL are found to be $5\times10^{19}$ cm$^{-3}$ with an efficiency of 25.64\% and FF of 80.32\%.

\subsection{Defect density optimization}
\begin{figure}
    \centering
    \includegraphics[scale=.7]{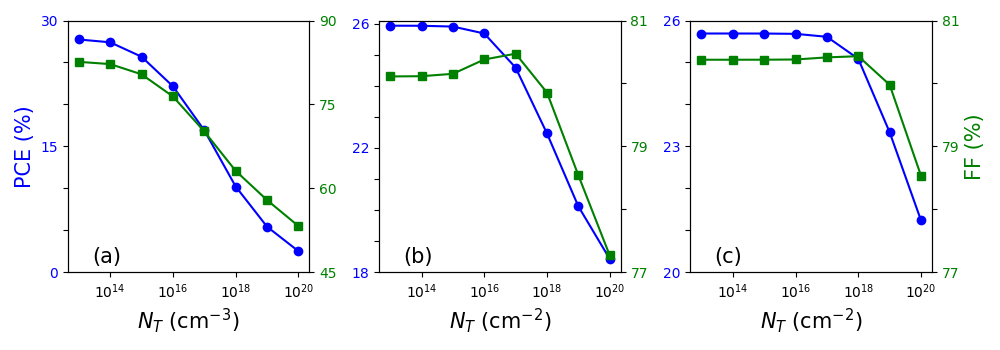}

    \caption{Variation of photo conversion efficiency (PCE) and fill factor (FF) with defect densities ($N_T$) in different layers: (a) AL, (b) AL/HTL interface and (c) ETL/AL interface.}
    \label{nt}
\end{figure}

The performance of PSCs is notably influenced by defects within the perovskite layer, including vacancies like  interstitials, Schottky and Frenkel defects. High defect concentrations significantly degrade the stability of PSCs \cite{Rai2020}. To examine the impact of defect density on device performance, we varied the absorber defect density from $10^{13} $ cm$^{-3}$ to $10^{20} $ cm$^{-3}$. Figure \ref{nt}(a) depicts how the PCE varies with defect density of AL, where the device parameters decline rapidly after $10^{15} $ cm$^{-3}$. This performance degradation is attributed to the increased presence of traps and recombination pathways \cite{Johnston2016}. Hence the threshold defect density of $10^{15} $ cm$^{-3}$ is considered for further optimization.

The defect density at the interfaces of AL with the HTL and ETL plays a crucial role in determining the performance of PSCs. Figure \ref{nt}(b) illustrates the variation of efficiency and FF with respect to the defect density at the AL/HTL interface. Both parameters remain stable at low defect densities up to approximately $10^{17}$ cm$^{-2}$. Beyond this point both efficiency and FF experience a sharp decline due to higher interface defect density. Similarly, figure \ref{nt}(c) shows the effect of defect density at the ETL/AL interface. While efficiency maintains relatively high value up to a defect density of around $10^{15}$ cm$^{-2}$ and FF having a maximum at $10^{17}$ cm$^{-2}$, a significant drop is observed at higher defect densities suggesting that charge extraction is severely hindered by interface defects. In expense of a slight less value of FF the defect density of $10^{15}$ cm$^{-2}$ is chosen for a higher efficiency. 
\begin{figure}[ht]
    \centering

    \includegraphics[scale=.9]{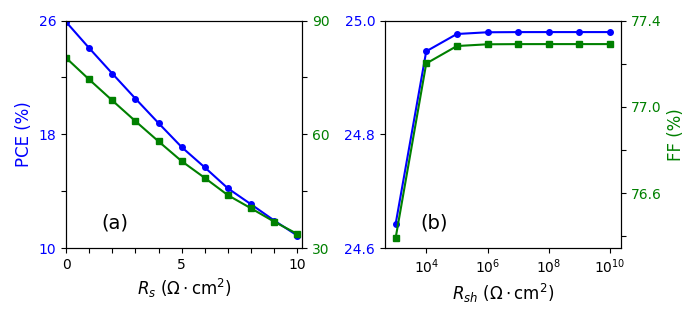}
   
    \caption{Variation of photo conversion efficiency (PCE) and fill factor (FF) with respect to (a) series ($R_s$) and (b) shunt ($R_{sh}$) resistance.}
    \label{r}
\end{figure}

\subsection{ Circuit resistance and back contact optimization}

\begin{figure}[h!]
    \centering
    \includegraphics[scale=0.8]{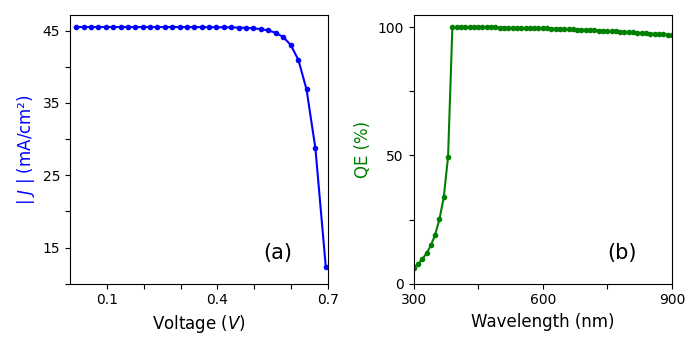}
        \caption{(a) Total current density ($\vert J \vert$) vs voltage ($V$) characteristic and (b) quantum efficiency vs wavelength plot for final optimized configuration FTO/SnO$_2$/RbGeI$_3$/CuI/Au.}
        \label{qe}
\end{figure}

 We have explored the dependency of efficiency and FF on the series resistance, $R_s$. As shown in Figure \ref{r}(a), the efficiency of the PSC decreases significantly with increase in $R_s$. For optimal performance, a series resistance of 0.5 $\Omega$·cm$^2$ is selected. To further understand the impact of circuit parameters on device performance, we have also examined the influence of shunt resistance, $R_{sh}$, on PSC efficiency. As shown in figure \ref{r}(b), increasing $R_{sh}$ from 10$^3$ $\Omega$·cm$^2$ to 10$^10$ $\Omega$·cm$^2$ resulted in the increase in efficiency and FF. The saturation achieved at $R_{sh}$ = $10^5$  $\Omega$·cm$^2$, hence chosen to be the optimal choice with an efficiency of 24.97\% . 
The back contact material is initially configured with Ag, possessing a work function of 4.57 eV  \cite{ef}. To optimize the performance, we have examined with materials such as Cu (4.65 eV) \cite{ef}, Au (5.1 eV) \cite{ef}, Ni (5.5 eV) \cite{back}, Pt (5.7 eV) \cite{back} and Fe (4.81 eV) \cite{back}. Back contact Ag shows a PCE of 24.97\% and FF of 77.28\%, while Cu shows a PCE of 25.71\% and FF of 79.66\% and all others exhibit the efficiency of 25.76\% with FF 79.81\%. Among them gold (Au) is the most used material as a back contact for its chemical inertness, corrosion resistance and high electrical conductivity \cite{Jaiswal2023}. Hence Au is considered to be the back contact metal for the PSC. The final configuration of the PSC is FTO/SnO$_2$/RbGeI$_3$/CuI/Au, with all the optimized parameters, gives an efficiency of 25.76 \% , FF of 79.81 \%, $V_{OC}$ and $J_{SC}$ of 0.709 V and 45.50 mA/cm$^2$ respectively. Figure \ref{qe} demonstrates the total current density ($\vert$ J $\vert$) vs voltage (V) characteristic for the final configuration along with the quantum efficiency (QE) of the PSC for the wavelength range of 300 nm to 900 nm, which shows up to $\sim $ 100\% of QE in the visible range. 
    
\section{Conclusions}
We have studied the electronic and optical properties of the rhombohedral phase of germanium-based rubidium halides, RbGeX$_3$ (X = Cl, Br, I). Various electronic properties have been explored, such as the band gap through electronic band structure calculations and the contributions of different orbitals via partial density of states calculations. From the study of electronic properties, we conclude that RbGeX$_3$ for X = Cl, Br, I have direct band gaps of 2.36 eV, 1.47 eV and 0.96 eV respectively by using the TB-mBJ potential with spin-orbit coupling, suggesting that the band gap decreases with the increasing ionic radius of the halide atoms. The lowest band gap is obtained for RbGeI$_3$ where the 4p orbital of Ge predominantly contributes to the VBM and the 5p orbital of iodine contributes to the CBM. Optical properties such as the dielectric function, refractive index, absorption coefficient are studied through first-principles calculations. For X = Cl, Br and I, the Static dielectric constants are found to be 3.98, 5.16 and 7.14 in ($xx$, $yy$) direction and 3.82, 5.15, 6.99 along $zz$, while the static refractive indices are 1.99, 2.27 and 2.67 along ($xx$, $yy$) and 1.95, 2.27, 2.64 along $zz$, respectively, showing an increasing trend from Cl to I and anisotropy along the $zz$ direction for all three halides. From the comparison of germanium-based rubidium halides in rhombohedral phase, RbGeI$_3$ is found to be the most promising candidate for photovoltaic applications. Hence, using RbGeI$_3$ as the absorber layer, we have constructed an all-inorganic PSC using SCAPS-1D. By studying various inorganic transport layers and optimizing thicknesses, doping densities, defect densities, circuit resistances and the back contact, we achieved an optimal efficiency of 25.76\% with a fill factor of 79.81\% for the FTO/SnO$_2$/RbGeI$_3$/CuI/Au configuration. This work motivates experimental exploration of R3m-RbGeI$_3$ based perovskite solar cells and offers insights into all-inorganic, non-toxic germanium based solar technologies for sustainable energy solutions.

\section*{Acknowledgments}
Authors acknowledge the computing resource `PARAM SHAVAK' at Computational Condensed Matter Physics lab, Department of Physics, NIT Silchar. The SCAPS-1D program was kindly provided by Dr. M.
Burgelman of the University of Gent in Belgium.
\bibliographystyle{unsrt}
\bibliography{main}
\end{document}